\documentclass[fleqn,twoside]{article}%
\usepackage{amsmath}
\usepackage{amsfonts}
\usepackage{amssymb}
\usepackage{graphicx}
\usepackage{bm}
\usepackage[utf8]{inputenc}
\usepackage[english]{babel}
\usepackage{multirow}
\usepackage{cite}

\newcommand{\be}{\begin{equation}}
\newcommand{\ee}{\end{equation}}
\newcommand{\bey}{\begin{eqnarray}}
\newcommand{\eey}{\end{eqnarray}}
\def\bes{\begin{equation}\begin{split}&}
\newcommand{\bi}{\bibitem}

\title  {A viable form of the metric Teleparallel F(T) theory of gravity.}
\author{Manas Chakrabortty$^1$, Nayem Sk$^{2}$ and Abhik Kumar Sanyal$^3$.\\
~~~~~~~~~~~~~~~\\
$^1$Dept. of Physics, Bankura University, Bankura, India - 722155\\
$^2$Dept. of Physics, Saidabad Manindra Chandra Vidyapith, Murshidabad, India - 742103\\
$^3$ Dept. of Physics, Jangipur College, Murshidabad, India - 742213}
\begin{document}
\maketitle
\footnote{
Electronic address:\\
$^{1}$manas.chakrabortty001@gmail.com\\
$^{2}$nayemsk1981@gmail.com\\
$^{3}$sanyal$_{-}$ak@yahoo.com\\}
\begin{abstract}

\noindent
Unlike $F(R)$ gravity, pure metric $F(T)$ gravity in the vacuum dominated era, ends up with an imaginary action and is therefore not feasible. This eerie situation may only be circumvented by associating a scalar field, which can also drive inflation in the very early universe. We show that, despite diverse claims, $F(T)$ theory admits Noether symmetry only in the pressure-less dust era in the form $F(T) \propto T^n, ~n$ being odd integers. A suitable form of $F(T)$, admitting a viable Friedmann-like radiation dominated era, together with early deceleration and late-time accelerated expansion in the pressure-less dust era, has been proposed.\\
\end{abstract}
\maketitle

\section{Introduction}

It is almost unanimously believed that the luminosity distance versus redshift (non-linear) curve observed from Sn1a data, is a consequence of recent accelerated expansion of the universe. A wide class of scalar field (dark energy) theories were promoted to explain the fact. Consequently, search for dark energy in the laboratory was initiated almost a decade back \cite {1.01, 1.02}, and recently it has apparently been ruled out following a laboratory based atom interferometry experiment, performed with incredible precision \cite{1.03}. As an alternative to the dark energy, initially, modified theories of gravity, such as $F(R), F(G), F(R,G)$ etc., were developed and extensively studied. Due to the fact that these theories suffer from Ostrogradsky instability, lately, alternative theories of gravity (metric and symmetric teleparallel theories) are in the lime light. These are called `alternatives', since instead of curvature, which is the building block of `general theory of relativity' (GTR), metric and symmetric teleparallel theories are built from the torsion and non-metricity (general affine connection) respectively. Our concern in this manuscript, is the generalised version ($F(T)$) of metric teleparallel gravity theory.\\

In the recent years, a generalized version of the so-called `teleparallel gravity' \cite{1.04}, namely the $F(T)$ theory of gravity (where, $T$ stands for the trace of the torsion tensor), also dubbed as `gravity with torsion' has been proposed as an alternative to both the dark energy issue and the modified theories of gravity. Factually, $F(T)$  teleparallel theory of gravity was primarily proposed to drive inflation \cite{1.1, 1.2}. Later, it was applied to drive the current accelerated expansion of the present universe without considering dark energy \cite{1.3, 1.4}. Thereafter, it has drawn a lot of attention and has been explored vastly in different contexts \cite{1.5,1.6,1.7,1.8,1.9,1.10,1.11,1.12,1.13,1.14,1.15,1.16,1.17,1.18,1.19,1.20,1.21,1.22,1.23,1.24,1.25,1.26,1.27,1.28,1.29,1.30,1.31,1.32,1.33,1.34,
1.35,1.35a,1.35b,1.35c}. It is worth mentioning that $F(T) \propto T$ leads to GTR, apart from a divergent term, and therefore, it is not required to supplement the action with the Gibbons-Hawking-York term. The very interesting feature of $F(T)$ gravity is that, very much like Lanczos-Lovelock gravity, it gives second-order field equations, as a result of which Ostrogradsky's instability is avoided. In this manuscript, our aim is to explore a physically reasonable form of the metric teleparallel theory. For the sake of self-standing, we briefly demonstrate the basic ingredients of the metric teleparallel theory and review the applicability of Noether theorem in different contexts.\\

The modified teleparallel action of $F(T)$ gravity is given by the following action,

\be\label{1.1}\mathbb{A} = \int  d^4 x  \mid e \mid  F(T)+ S_m ,\ee
in the units $c = 16 \pi G = 1$, where $|e| = \det {e^{i}_{\mu}}=\sqrt {-g}$ and $S_m$ is the matter action. Teleparallelism \cite{1.36} uses a vierbein field $\mathbf{e}_{i} (x^{\mu})$ as dynamical object (it consists of four linearly independent vector fields forming a local basis for the tangent bundle, instead of the coordinate basis), which is an orthonormal basis for the tangent space at each point $x^{\mu}$ of the manifold $\mathbf{e}_{i}$, where $\mathbf{e}_{ij}={\eta}_{ij}$, $i$ runs from $0, 1, 2, 3$, and ${\eta}_{ij} = \mathrm{diag}(1,-1,-1,-1)$. The vector $\mathbf{e}_{i}$ can be described by its components $e^{\mu}_{i}$, $\mu$ = 0, 1, 2, 3 in a coordinate basis, i.e. $\mathbf{e}_{i}=e^{\mu}_{i}\partial\mu$. In the above, Latin indices refer to the tangent space, while Greek indices label coordinates on the manifold. The metric tensor is obtained from the dual vierbein as $g_{\mu\nu}(x)=\eta_{ij}e^{i}_{\mu}(x) e^{j}_{\nu}(x)$. In contrast to the (GTR) which uses the torsion-less Levi-Civita connection, as already mentioned, the curvature-less Weitzenb$\ddot{\mathrm{o}}$ck connection is applied in `Teleparallelism' \cite{1.37}, for which the non-vanishing torsion is,

\be\label{1.2} T^{\lambda}_{\mu\nu} \equiv e^{\lambda}_{i}[\partial_{\mu}e^{i}_{\nu}-\partial_{\nu}e^{i}_{\mu}].\ee
The above tensor encompasses all the information regarding the gravitational field. The teleparallel equivalent of General Relativity (TEGR) Lagrangian is built with the torsion \eqref{1.2}, and its dynamical equations for the vierbein imply Einstein's equations for the metric. The teleparallel Lagrangian given in \cite{1.38,1.39,1.40} is,

\be\label{1.3} L_T = {S_{\rho}}^{\mu\nu} {T^{\rho}}_{\mu\nu},\ee
where, the tensor ${S_{\rho}}^{\mu\nu}$

\be\label{1.4} {S_{\rho}}^{\mu\nu} =  \frac{1}{2}[{K^{\mu\nu}}_{\rho}+{\delta}^{\mu}_{\rho}{T^{\theta\nu}}_ {\theta}-{\delta}^{\nu}_{\rho}{T^{\theta\mu}}_{\theta}].\ee
is built from the contorsion tensor, $K^{\alpha\beta}_{\rho}$, given by

\be\label{1.5} K^{\mu\nu}_{ \rho} =  -\frac{1}{2}[{T^{\mu\nu}}_{\rho}-{T^{\nu\mu}}_ {\rho}-{T_{\rho}}^{\mu\nu}],\ee
which equals the difference between Weitzenb$\ddot{\mathrm{o}}$ck and Levi-Civita connections.\\

Now, in order to study cosmological consequence of the said alternative theory of gravity, a particular form of $F(T)$ is required. Before setting a form of $F(T)$ by hand or reconstructing it from the late history of cosmic evolution, it is always desirable to find its form following some physical consideration; as for example, invoking Noether symmetry, which had been applied earlier in different modified theories of gravity in particular, with tremendous success. De Ritis et.al. \cite{1.41} had applied Noether symmetry for the first time in scalar-tensor theory of gravity, and found an exponential form of the potential, which can drive inflation in the very early universe. Such an exciting result inspired authors to extensively apply Noether symmetry in different theories of gravity, such as, scalar-tensor theories, higher order theories, Gauss-Bonnet gravity, quantum cosmology and also $F(R)$ theory, and so on. Particularly, since $F(T)$ gravity appeared as an alternative to $F(R)$ gravity theory, therefore, for the sake of comparison, we enlist in table 1, the available Noether symmetries for $F(R)$ theory of gravity, in the background of isotropic and homogeneous Robertson-Walker (RW) metric

\be\label{1.6} {ds}^2 = - {dt}^2 + {a^2(t)}\left[\frac{dr^2}{1-kr^2} + r^2 \big(d\theta ^2 + \sin^2{\theta}~d\phi^2\big)\right],\ee
in different eras \cite{1.71, 1.72, 1.73, 1.74, 1.75}. Likewise, application of Noether symmetry to explore suitable forms of $F(T)$ is also available in the literature \cite{1.80, 1.81, 1.82, 1.83, 1.84, 1.85}. However, due to diffeomorphic invariance, the $(^0_0)$ and $(^i_0)$ equations of Einstein are altogether four: the energy (Hamiltonian) and the momenta constraint equations. Thus, Gravity is a constrained system, and Noether symmetry is not on-shell for such a constrained system, and every available Noether symmetry of gravitational theory, is required to satisfy the four aforesaid equations \cite{1.86,1.87,1.88}. Alternatively, application of (modified) Poisson first theorem happens to be a straight forward technique to check consistency, in this regard \cite{1.89, 1.89a}. In $F(T)$ gravity theory however, apart from the metric coefficients, the configuration space is spanned by ($T,~\dot T$). Unfortunately, due to absence of $\dot T$ term in the action, the Hessian determinant vanishes and the action becomes singular. As a result, the phase-space structure remains obscure \cite{1.90}, and Poisson theorem cannot be applied. Hamiltonian has been constructed following Dirac constrained analysis, but it contains momenta in the denominator and hence impossible to handle \cite{1.90}. In the context of RW metric, it is therefore suggestive either to incorporate lapse function ($N$) in the symmetry equation, without fixing it a priori \cite{1.91,1.92}, or to incorporate the constraint in the Noether equation, through a Lagrange multiplier \cite{1.93,1.94}. However, in both the situations, the lapse function and the Lagrange multiplier remain arbitrary, and one does not know, which form would give a symmetry. \\

\begin{figure}
    \begin{minipage}[h]{0.90\textwidth}
    \scalebox{0.95}{
  \begin{tabular}{|c|c|c|c|c|c|}
   \hline
   \multicolumn{2}{|c|}{Vacuum dominated era}  & \multicolumn{2}{c|}{Radiation dominated era} & \multicolumn{2}{c|}{Matter dominated era} \\
   \hline
   Form of $F(R)$ & Conserved currents & Form of $F(R)$ & Conserved currents & Form of $F(R)$ & Conserved currents \\
   \hline
   & & & & &  \\
   $F_0 R^2$ & $a^3 \dot R$ & & & &  \\
   & & & & &  \\
   $F_0 R^{3\over 2}$ & ${d\over dt}(a\sqrt R)$ & & & & \\
   & & $F_0 R^2$ & $a^3 \dot R$ & $F_0 R^{3\over 2}$ & ${d\over dt}(a\sqrt R)$ \\
   $F_0 R^{7\over 5}$ & $\sqrt{a}{d\over dt}(aR^{2\over 5})$ & & & & \\
   & & & & &  \\
   $F_0 R^{-1}$ & $\sqrt{a}\dot aR$ & & & & \\
   \hline
\end{tabular}
}
\caption{Available Noether symmetries in RW space-time for $F(R)$ theory of gravity in different eras.}
\label{tab:Table 1}
\end{minipage}%
 \end{figure}
In the literature, although a host of Noether symmetries for $F(T)$ gravity, together with associated conserved currents had been explored, \cite{1.80, 1.81, 1.82, 1.83, 1.84, 1.85}, the consistency of the available Noether symmetries in connection with the ($^0_0$) equation of Einstein, had not been inspected. We therefore take up this task in the present manuscript. In the following section, we explore Noether symmetry following the Lagrange multiplier technique for $F(T)$ teleparallel gravity, in the background of spatially flat homogeneous and isotropic Robertson-Walker space-time. We also incorporate the available symmetry in the energy ($^0_0$) equation of Einstein, for a consistency check. It is unveiled that pure $F(T)$ leads to imaginary action in the vacuum dominated era, and the only propitious result found is $F(T) \propto T^n$, but for odd integral values of $n$, in the pressure-less dust era. In section 3, we try to construct the form of $F(T)$ in view of a viable (Friedmann-like decelerated) radiation era. In view of all our findings we propose a reasonable form of $F(T)$ in section 4. Section 5 concludes our work.

\section{Noether Symmetry analysis in F(T) Teleparallel Gravity:}

Before we commence, let us enlist all the claims \cite{1.80}-\cite{1.85} for available forms of $F(T)$ together with associated conserved currents in different eras in table 2. As already mentioned, primarily our aim is to scrutinize such claims. For this purpose, our starting point is action \eqref{1.1}, in which the matter action $S_m$ includes both perfect fluid and dark matter associated with a barotropic equation of state $\omega$. In the spatially flat Robertson-Walker space-time,

\be\label{2.1} {ds}^2 = - {dt}^2 + {a^2(t)}\big[dr^2 + r^2 \big(d\theta ^2 + \sin^2{\theta}~d\phi^2\big)\big],\ee
the vierbein is
\be\label{2.2}e^{i}_{\mu}= \mathrm{diag}(1,a(t),a(t),a(t)).\ee
In the above, $a(t)$ is the cosmological scale factor. Our motive is to explore Noether symmetry in different eras, satisfying the only constraint, viz., the $(^0_0)$ equation of Einstein.\\

\begin{figure}[t!]
    \begin{minipage}[h]{0.90\textwidth}
    \scalebox{0.95}{
  \begin{tabular}{|c|c|c|c|c|}\hline
 \multirow{2}{*}{Citations} & \multicolumn{2}{|c|}{Vacuum dominated era} & \multicolumn{2}{c|}{Matter dominated era} \\ \cline{2-5}
 & Form of $F(T)$ & Conserved currents & Form of $F(T)$ & Conserved currents \\ \cline{1-5}
 \multirow{3}{*}{[50]} & & & &\\
 & $F_0 T^{(-\frac{3}{C})}$ & $36\frac{a^{(\frac{C}{2}+2)}\dot a T^{-(1+\frac{3}{C})}}{C}$ & & \\
 & & & &\\\cline{1-3}
 \multirow{6}{*}{[51]} & & & &\\
 & \multirow{4}{*}{$F_0 T^n$} & $(2n-1)t a^3 T^n  + 4n (n-1) a^2  \dot a T^{n-1}$ & &\\\cline{3-3}
 & & & &\\
 & & $(2n-1) a^3 T^n$ & &\\\cline{3-3}
 & & & &\\
 & & $-4n^2 a^{(2-\frac{3}{2n})} \dot a T^{n-1}$ & &\\\cline{1-3}
 \multirow{7}{*}{[52]} & & & &\\
 & $F_0 T^n$ $(n\neq \frac{3}{2}, \frac{1}{2})$ & $\left(\frac{3C}{2n-1} t\right)H-12F_0 n\left(C a^2+c_3 a^{(2-\frac{3}{2n})}\right) \dot a T^{n-1}$
 & & \\
 & & & &\\\cline{2-3}
 & & & &\\
 & $F_0 T^{3/2}$ & $\frac{1}{5}(3c-2m)t H - 18 F_0 \left[\left(\frac{c}{2}-\frac{m}{6}\right) a^2+c_4 a\right] \dot a T^{1/2}$ & &\\
 & & & &\\\cline{2-3}
 & $F_0 T^{1/2}$ & $c_1 t E-6 F_0 \left(-2c_1 a +c_3 a^\frac{3}{4}\right) \dot a T^{-\frac{1}{2}}$ & &\\ \cline{1-5}
 \multirow{2}{*}{[53]} & & & &\\
 & & & $F_0 T^n$ & $-12 \alpha_0 a^{(2-\frac{3}{2n})} \dot a F'(T)$\\ \cline{1-1}
 \cline{4-5}
 \multirow{2}{*}{[54]} & & & &\\
  & & & $F_0 T^n$ & $-12 \alpha_0 a^{(2-\frac{3}{2n})} \dot a F'(T)$\\ \cline{1-1} \cline{4-5}
  \multirow{2}{*}{[55]} & & & &\\
 & & & $F_0 T^{3/2}$ & $a \dot a F' (T)$\\ \cline{1-5}
\end{tabular}
}
\caption{Claimed Noether symmetries in RW metric for $F(T)$ theory of gravity in different eras.}
\label{tab:Table 2}
\end{minipage}%
 \end{figure}

Canonical formulation of $F(T)$ theory following Lagrange multiplier technique \cite{1.80,1.81,1.82,1.83,1.84,1.85} may be cast with finite degrees of freedom only. In this formalism, $T + 6 {\dot a^2\over a^2}=0$ is treated as a constraint in the spatially flat Robertson-Walker metric \eqref{2.1}. This constraint is introduced in the action (\ref{1.1}) through a Lagrange multiplier $\lambda$. In the presence of a barotropic fluid, the action may therefore be expressed as,

\bes\label{2.3}\mathbb{A} = 2{\pi}^2\int\Big[F(T) - \lambda\Big\{T + 6\Big({\dot a^2\over a^2} \Big)\Big\}-\rho_{0} a^{-{3(\omega+1)}}\Big]a^3 dt ,\end{split}\ee
where, $\rho_0$ is the present value of matter density, either for radiation ($\rho_{r0}$) or pressure-less dust ($\rho_{m0}$), and $\omega$ is the equation of state parameter. Now varying the action with respect to $T$ one gets $\lambda = F'(T)$, where $F'(T)$ is the derivative of $F(T)$ with respect to $T$. Substituting the expression of $\lambda$, one can express the above action \eqref{2.3} in the following form,

\bes\label{2.4} \mathbb{A} = 2{\pi}^2\int[-6a \dot a^2 F'+ a^3(F-F'T)- \rho_{0} a^{-3\omega}]dt.\end{split}\ee
It is important to mention that, being devoid of the time derivative of $T$, the Hessian determinant vanishes. Therefore the above action, unlike $F(R)$ theory of gravity, is singular \cite{1.90}, although it is often referred to as canonical. Nonetheless, for Noether symmetry analysis, such an action is well-suited. The point Lagrangian is therefore,

\be\label{2.5} L(a,\dot a,T ,\dot T) = \left[-6a \dot a^2 F'+ a^3(F-F'T)- \rho_{0} a^{-3\omega}\right],\ee
and the field equations in terms of the Hubble parameter $H = {\dot a\over a}$ are,

\be\label{2.6} 48H^2\dot{H}F'' - 4(\dot{H}+3H^2)F'-F -\omega\rho_{0} a^{-{3(\omega+1)}}= 0\ee
\be\label{2.7} T = -6H^2\ee
\be\label{2.8} F - 2TF'-\rho_{0} a^{-{3(\omega+1)}}=F +12H^2F' - \rho_{0} a^{-{3(\omega+1)}}=0\ee
Note that, the constraint has been retrieved in equation \eqref{2.7}. One can also trivially check that the field equations of GTR may be recovered for $F(T) \propto T = -6H^2$. As mentioned, we explore Noether symmetries corresponding to the point Lagrangian \eqref{2.5} in the following . Noether theorem states that, if there exists a vector field $X$, for which the Lie derivative of a given Lagrangian $L$ vanishes i.e. $\pounds_X L =X L = 0 $, the Lagrangian admits Noether Symmetry together with a conserved current $\Sigma = i_X \theta_L$, where $\theta_L$ is the Cartan one form. We repeat that Noether symmetry is usually on-shell, but not for constrained system, we are dealing with. So once we find a symmetry, we shall check its consistency in connection with equation \eqref{2.8}.\\

\noindent
\textbf{Noether equations:} The configuration space of the Lagrangian \eqref{2.5} under consideration is $M(a,T)$ and the corresponding tangent space is $T_M(a,T,\dot a, \dot T)$. Hence the generic infinitesimal generator of the Noether Symmetry is,

\be\label{N1} X = \alpha_1 \frac{\partial }{\partial a}+\beta_1\frac{\partial }{\partial T} +\dot\alpha_1 \frac{\partial }{\partial\dot a}+ \dot\beta_1\frac{\partial }{\partial\dot T},\ee
where, $\alpha_1 = \alpha_1(a,T)$ and $\beta_1 = \beta_1(a,T)$. Further, the Cartan one form is:
\be \label{cartan1}\theta_{L} = {\partial L\over \partial a} da + {\partial L\over \partial T} dT,\ee
and the associated constant of motion is, $\Sigma = i_{X} \theta_{L}$, as already mentioned. Now using the existence condition for Noether Symmetry, viz.  $\pounds_{X} L = X L = 0$, we obtain the Noether equation. Thereafter we equate the coefficients of $\dot{a}^2$, $\dot{T}^2$, $\dot a \dot T$ along with the term free from derivatives respectively to zero as usual, and obtain the following system of partial differential equations:

\be\label{N2} \alpha_1 F' + \beta_1 a F''+ 2a F'\alpha_{1,a}= 0,\ee
\be\label{N3} \alpha_1' = 0,\ee
\be\label{N4} 3\alpha_1 \left({F-T F'}\right)+3\omega\rho_{0} a^{-{3(\omega+1)}}-a\beta_1 T F'' =0.\ee

\subsection{Available symmetries in pure vacuum era:}

It is important to mention that, particles are created, once inflation is over, due to the oscillation of the scalar field. Thus, very early vacuum-dominated era might contain a scalar field, but is devoid of barotropic fluid in any of its form. In the absence of the scalar field and for $\rho = 0 = p$, the field equations \eqref{2.6} and \eqref{2.8} are considerably simplified, and combined to yield

\be\label{N8.1}48H^2\dot{H} F''-4\dot{H} F' = 0,\ee
which may immediately be solved to obtain $F(T) \propto T^{\frac{1}{2}}$, which is the only allowed form of $F(T)$ theory in vacuum. It is not clear, despite such unique form, why attempts to find forms of $F(T)$ were taken up. In fact, authors \cite{1.80, 1.81, 1.82} attempted to find Noether Symmetry in pure vacuum era and ended up with $F \propto T^n$. Let us now examine why such uncanny results appeared. The above set of Noether equations \eqref{N2}, \eqref{N3} and \eqref{N4} admit the following solutions in pure vacuum era, viz.
\be\label{N5} \alpha_1= a^{1-\frac{3}{2n}}, \;\;\;\beta_1 = -\frac{3}{n} {a}^{-\frac{3}{2n}} T,\;\;\; F(T) = F_0 T^{n},\ee
and its associated conserved current, which reads as,
\be\label{N6} \Sigma = -12 a^{(2-\frac{3}{2n})}\dot a F'(T)=-12 n F_0 a^{(2-\frac{3}{2n})}\dot aT^{n-1}.\ee
In the above and everywhere else, $F_0 > 0$, ensures retrieval of GTR ($n = 1$). One can now immediately solve the above equation for $a(t)$ as,
\be \label{N7} a(t) = {\left(\frac{3}{2n}\right)}^\frac{2n}{3} {\left[\left(-\frac{1}{6}\right)^{n-1}\left(-\frac{\Sigma_1}{12 n F_0} \right)\right]}^\frac{2n}{3(2n-1)} {(t-t_0)}^\frac{2n}{3}.\ee
Although, the form of $F(T) \propto T^n$ \eqref{N5} so obtained via Noether symmetry analysis, in the very early vacuum dominated era, does not admit de-Sitter solution, the solution of the scale factor \eqref{N7} is clearly found to admit power law inflation for $n > {3\over 2}$. For example, $n = 3$ implies $a(t)\propto (t-t_0)^2$, and $n > 3$ gives even faster rate of the expansion of the early universe. However, before coming to a conclusion, note that the ($^0_0$) equation \eqref{2.8} for $F(T)=F_0 T^n$ in vacuum dominated era takes the form,

\be\label{N8}E_L=-a^3 \left(F - 2TF'\right)=(2n-1)F_0 a^3 T^n = 0,\ee
and the above equation is satisfied for none other than $n = {1\over 2}$. Now for $n=\frac{1}{2}$ the scale factor $a(t)\propto {(t-t_0)}^\frac{1}{3}$. As a result only decelerated expansion is administered in the very early universe.
Clearly the claim that $F(T) \propto T^n$, for arbitrary $n$ \cite{1.80, 1.81, 1.82} is patently false. Note that, such a form of $F(T)$ is meaningless, since $F(T) \propto T^{1\over 2} = i\sqrt{6} \left({\dot a\over a}\right)$ turns the action, imaginary. On the contrary $F(R)$ theory of gravity, admits at least four reasonably viable forms along with their associated conserved currents as depicted in table-1. Undoubtedly, vacuum era of $F(R)$ gravity has much reacher structure than $F(T)$ teleparallel theory of gravity. It may be mentioned that in the presence of a scalar field on the contrary, Noether symmetry remains obscure, and arbitrary form of $F(T)$ is admissible. It is therefore suggestive to incorporate a scalar field (dilatonic or Higgs field) which would be responsible to drive inflation in the action $F(T)$ \cite{1.94a}.

\subsection{Available symmetries in radiation-dominated era:}

In the radiation dominated era, Noether symmetry does not exist, and indeed there is no such claim in the literature too, in this regard. However, in $F(R)$ theory of gravity, $F(R) = F_0 R^2$ admits Noether symmetry being associated with the conserved current $\Sigma = a^3 \dot R$ (table-1). Again we find that $F(R)$ gravity has richer structure than $F(T)$ teleparallel gravity theory in radiation dominated era.

\subsection{Available symmetries in pressureless dust era:}

In pressureless dust era, the above Noether equations \eqref{N2}, \eqref{N3} and \eqref{N4} significantly admit the same solutions \eqref{N5}, \eqref{N6} as already found in vacuum era. Clearly, the scale factor also admits the same solution \eqref{N7}. Apparently, the same solutions are found as already presented in \cite{1.83, 1.84}. But, we need to scrutinize the symmetries in the light of the energy constraint equation of Einstein. For the available symmetry $F(T)=F_0 T^n$, the $(^0_0)$ equation \eqref{2.8} takes the following form in the matter dominated era,

\be \label{N12}E_L = -a^3 \left(F - 2TF'-\rho_{m0} a^{-3}\right)=(2n-1)F_0 a^3 T^n+\rho_{m0}=0,\ee
which may again be solved for $a(t)$ to find,

\be \label{N13} a(t) ={\left(\frac{3}{2n}\right)}^\frac{2n}{3} {\left[\left(-\frac{1}{6}\right)^{n}\left(-\frac{\rho_{m0}}{(2n-1)F_0}\right)\right]}^\frac{1}{3} {(t-t_0)}^\frac{2n}{3}.\ee
Comparing the scale factors found in view of Noether symmetry analysis \eqref{N7} and the constraint $E_L = 0$ \eqref{N13}, it is possible to express $\rho_{mo}$ as,
\be \label{N14}\rho_{mo} =-(2n-1)F_0 {\left(-6\right)}^{\frac{n}{2n-1}} {\left[-\frac{\Sigma_1}{12 nF_0}\right]}^\frac{2n}{(2n-1)}=\left(-1\right)^\frac{3n-1}{2n-1} (2n-1)F_0 {\left[\frac{{\Sigma_1}^2}{24 n^2 {F_0}^2}\right]}^\frac{n}{(2n-1)}.\ee
The above expression of $\rho_{m0}$, gives rise to some important consequences. Firstly, $n \ne {1\over 2}$, so that some amount of matter $\rho_{m0}$ (the structures) remain present in the universe. Note that in the process, the pathological form $F(T) \propto \sqrt T = i\sqrt 6 \left({\dot a\over a}\right)$ is averted. Next, since $F_0 > 0$, as already mentioned, so to ensure positivity of $\rho_{m0}$, we must have $n > {1\over 2}$ and $3n-1 = 2m$, where $m > 0$ is an integer ($m = 0$ leads to $n = {1\over 3}$, which makes $F(T)$ imaginary, and hence is excluded). As a consequence, indeed one finds admissible Noether symmetry in the form $F(T) \propto T^n$, but only for the odd integral values of $n$, such as, $n = 1, 3, 5 ....$ etc. This indicates, apart from the Friedmann solution $\big[a(t) \propto t^{2\over 3}\big]$, corresponding to $F(T) \propto T, ~\Sigma \propto \sqrt a \dot a$; accelerated expansion, such as $a(t) \propto t^2$, corresponding to $F(T) \propto T^3,~\Sigma \propto a^5H^5$; $a(t) \propto t^{10\over 3}$ corresponding to $F(T) \propto T^5,~\Sigma \propto a^{27\over 10}H^9$ etc., are possible. This result is definitely encouraging. \\

\section{Reconstruction from radiation era}

The success of the standard (FLRW) model of cosmology lies in the radiation and the early matter (pressure-less dust) dominated era. Once the seeds of perturbation are found (in view of an very early inflationary scenario), the Friedmann-like decelerated radiation era exactly formulates the formation of structures (stars, galaxies, clusters and superclusters), along with the CMBR which occurred at a redshift value $z \approx 1080$. It is therefore primarily required to associate a decelerated expansion in the radiation era, to envisage a viable evolution history of the universe. It may be worth recapitulating that a Friedmann-like decelerated expansion $a(t) \propto \sqrt t$, requires the Ricci scalar to vanish ($R = 0$). In the case of torsion, although $F(T) \propto T$ leads to GTR, nonetheless, $T =0$, leads to a static universe, and a Friedmann-like decelerated expansion is not obvious. To inspect the situation here, we combine the field equations \eqref{2.6} and \eqref{2.8} to find,

\be \label{rhop} \rho + p = \rho_0(1+\omega) a^{-3(\omega+1)} = {4\over 3} \rho_{r0} a^{-4} = 48H^2\dot H F'' - 4 \dot H F' = -4H F'' \dot T - 4 \dot H F' = -4{d\over dt}(H F'),\ee
where, we have substituted $\omega = {1\over 3}$ for radiation era, and $\rho_{r0}$ stands for the current value of the amount of radiation present in the universe. Now seeking a solution in the form $a = a_0 t^n$, where $a_0$ and $n$ are constants, one can compute, $H = {n\over t},~\dot H = -{n\over t^2},~T = -6H^2 = -6{n^2\over t^2}$. Hence upon integration, one finds,

\be F' = f_1{t\over n} + {\rho_{r0}\over 3a_0^4 (4n-1) t^{2(2n-1)}} = f_1{\sqrt{-6}\over \sqrt T} + {\rho_{r0}\over 3a_0^4(4n-1)n^{4n-1}(-6)^{2n-1}}T^{2n-1},\ee
where, $f_1$ is a constant of integration. Further integration yields,

\be\label{form} F(T) = 2f_1\sqrt{-6T} + {\rho_{r0}\over 6a_0^4(4n-1)n^{4n}(-6)^{2n-1}}T^{2n},\ee
apart from a constant of integration, which does not contribute to the field equations. Note that the first term is essentially a divergent term in the RW metric under consideration. Thus, we are left with the second term only. Under the choice $n = {1\over 2}$, the radiation era evolves exactly like the standard (FLRW) model, and the action reduces to that of GTR $(F(T) \propto T)$. This is actually what we are searching for. Note that, one can also consider other forms of $F(T)$, by choosing $n$ judiciously, keeping in mind that $F(T) > 0$, to recover GTR, $n < 1$ for decelerated expansion, $n \ne {1\over 4}$ to avert divergence, and finally the action has to be real. Satisfying all these conditions one can easily check that, for ${1\over 4} < n < 1$, the best option is to choose $n = {1\over 2}$, while for $n < {1\over 4}$, the decelerated expansion is too slow to produce CMB at the right epoch.

\section{Proposition:}

It may be mentioned that the Ricci scalar measures the difference of areas between a curved space and the flat space of a sphere (say), formed by the set of all points at a very small geodesic distance. From Einstein's equation it is found to be equal to the trace $\mathrm{T}$ of the energy momentum tensor $\mathrm{T}_{\mu\nu}$, i.e. $R = R_{\mu\nu}g^{\mu\nu} = \mathrm{T}{\mu\nu}g^{\mu\nu} = \mathrm{T}$. For electromagnetic field the trace vanishes and hence the Ricci scalar. As a result, in the radiation dominated era, which initiated soon after the hot big-bang, the scalar curvature vanish $(R = 0)$, to realize a Friedmann-like radiation era $a \propto \sqrt t$. It is also important to mention that structure has to formation computed from the seeds of perturbation generated during early inflationary era, is based on the standard model of cosmology, which requires $R = \mathrm{T} = 0$, in the middle. This implies, unlike the scale factor, which evolves from a very small to a large value with the cosmic expansion, the Ricci scalar does not evolve in a continuous manner from a very large to an insignificantly small value. However, a viable $F(R)$ theory of gravity is still presented in the form $F(R) = \alpha R + \beta R^2 + \gamma R^{-1}$ \cite{1.98}, or $F(R) = \alpha R + \beta R^2 + \gamma R^{3\over 2}$ \cite{1.99}, with the preoccupied assumption that $R$ evolves from a very large at the earlier epoch to an insignificantly small value at present in a continuous manner, so that $R^2$ dominated in the early universe, leading to inflation, $R$ in the middle, to establish standard model, and $R^{-1}$/ $R^{\frac{3}{2}}$ at present, to envisage accelerated expansion of the universe. Clearly, this leads to a conceptual problem. On the contrary, when torsion is attributed to gravity, usually, a form such as $F(T) = f_0 T + f_1 T^2 + \cdots$ is chosen to combat early deceleration in the Friedmann form $\left(a(t) \propto t^{2\over 3}\right)$ followed by late-time cosmic acceleration in the matter (pressure-less dust) dominated era. In view of our analysis, it is clear that $F(T) \propto T$ gives exactly Friedmann-like radiation dominated era $(a(t) \propto \sqrt t)$, and pressure-less dust dominated era $\left(a(t) \propto t^{2\over 3}\right)$. Therefore, it is not required to set $T = 0$, at any stage. This has a great conceptual advantage over modified theories of gravity.\\

In view of our current analysis, we find that a scalar field must be associated in $F(T)$ gravity theory, not only to drive inflation, but also to avert the pathological behaviour of pure $F(T)$ gravity in the very early vacuum-dominated era. Next, $F \propto T$, gives the standard model of cosmology, both in the radiation and early matter dominated era. Finally, in view of Noether analysis, instead of $T^2$, one should associate $T^3$ and higher odd integral powers in the action. That is, a viable form that might explain the cosmic evolutionary history may be in the following form, $F(T) = f_0 T + f_1 T^3 + \cdots$, and the action may be proposed as,

\be \label{A} A = \int \left[f_0 T + f_1 T^3 + \cdots + {1\over 2}\dot\phi^2 - V(\phi)\right]\sqrt{-g}d^4x,\ee
where, the scalar field drives inflation at the very early stage, and decayed to an insignificant value in the process of particle creation. Additionally, $T$ dominates in the middle to envisage the standard model, while odd-integral higher order terms are responsible for late-time cosmic acceleration.

\section{Concluding remarks:}

Noether symmetry has been extensively studied in the literature for teleparallel gravity with torsion $F(T)$, and several claims were made regarding the forms of $F(T)$ and associated conserved currents in vacuum and pressure-less dust era, as presented in table 2. Nonetheless, it is trivial to note that pure $F(T)$ gravity is not empowered with a meaningful form in vacuum. This is a major shortfall of gravity with torsion, and at least a scalar field is required to forestall $F(T)$ gravity theory from such bizarre situation. Hence, unlike $F(R)$ gravity, in which $R^2$ term can steer inflation in the early universe, the scalar field can only be responsible to drive inflation in $F(T)$ gravity theory. Obviously, unless, it can be shown that the scalar is almost completely used up in producing particles at the end of inflation, and the rest are redshifted away, it would be a responsibility to observe its trace in the present universe. Otherwise, the main objective to explain late-time acceleration without dark energy, would be in vain.\\

It is also observed that radiation era does not yield any Noether symmetry, and as such there were no such claim too. On the contrary, it is important to mention that, $F(R)$ theory admits symmetry in radiation era, which are enlisted in table 1. Result of the present analysis also reveals the fact that the radiation and the early pressure-less dust era is best described by $F(T) \propto T$. This is a non-trivial result, since in GTR, the Ricci scalar is proportional to the trace of the energy-momentum tensor, which vanishes for radiation and as a result Friedmann-like ($a \propto \sqrt t$) solution is admissible. Although, $F(T) \propto T$ is TEGR, however, vanishing trace of energy-momentum tensor does not enforce $T = 0$, but still, Friedmann-like cosmic evolution is admissible.\\

Finally, we find that Noether symmetry indeed exists in the pressure-less dust era in the form $F(T) \propto T^n$, but only for odd integral values of $n$. In view of all these findings, we put forward an action \eqref{A}, which might possibly foretell the cosmic evolutionary history of the universe. Although, it is apparent that $F(R)$ gravity has a much richer structure than the $F(T)$ gravity theory, note that, unlike the Ricci curvature scalar, it is not required to set torsion to vanish at any stage of cosmic evolution. Further, power lower than first degree in $R$ is required to explain late-time cosmic acceleration. Contrarily, in the case of torsion higher degree terms are required for the simple reason that, the Hubble parameter is reduced with expansion, and higher degree terms in Hubble parameter falls off even faster, reducing the torsion considerably, and eventually leading to late-time cosmic acceleration. In a nut-shell, late-time acceleration is an outcome of lesser torsion. In this sense, $F(T)$ gravity is apparently free from conceptual problem.\\

\end{document}